\documentclass[aip,amsmath,amssymb,graphicx]{revtex4-1}
\usepackage{graphicx}
\usepackage{dcolumn}
\usepackage{bm}

\draft 

\begin{document}


\title{Electron Transport in InAs-InAlAs Core-Shell Nanowires} 




\author{Gregory W. Holloway}
\thanks{these authors contributed equally}
\altaffiliation{Department of Physics and Astronomy, University of Waterloo, Waterloo, Ontario, N2L 3G1, Canada}
\altaffiliation{Waterloo Institute for Nanotechnology, University of Waterloo, Waterloo, Ontario, N2L 3G1, Canada} 
\affiliation{Institute for Quantum Computing, University of Waterloo, Waterloo, Ontario, N2L 3G1, Canada}

\author{Yipu Song}
\thanks{these authors contributed equally}
\altaffiliation{Department of Chemistry, University of Waterloo, Waterloo, Ontario, N2L 3G1, Canada}
\affiliation{Institute for Quantum Computing, University of Waterloo, Waterloo, Ontario, N2L 3G1, Canada}

\author{Chris M. Haapamaki}
\affiliation{Department of Engineering Physics, Centre for Emerging Device Technologies, McMaster University, Hamilton, ON, L8S 4L7, Canada}

\author{Ray R. LaPierre}
\affiliation{Department of Engineering Physics, Centre for Emerging Device Technologies, McMaster University, Hamilton, ON, L8S 4L7, Canada}

\author{Jonathan Baugh}
\email{baugh@iqc.ca}
\altaffiliation{Department of Chemistry, University of Waterloo, Waterloo, Ontario, N2L 3G1, Canada}
\altaffiliation{Department of Physics and Astronomy, University of Waterloo, Waterloo, Ontario, N2L 3G1, Canada}
\altaffiliation{Waterloo Institute for Nanotechnology, University of Waterloo, Waterloo, Ontario, N2L 3G1, Canada} 
\affiliation{Institute for Quantum Computing, University of Waterloo, Waterloo, Ontario, N2L 3G1, Canada}

\date{\today}

\begin{abstract}
Evidence is given for the effectiveness of InAs surface passivation by the growth of an epitaxial In$_{0.8}$Al$_{0.2}$As shell. The electron mobility is measured as a function of temperature for both core-shell and unpassivated nanowires, with the core-shell nanowires showing a monotonic increase in mobility as temperature is lowered, in contrast to a turnover in mobility seen for the unpassivated nanowires. We argue that this signifies a reduction in low temperature ionized impurity scattering for the passivated nanowires, implying a reduction in surface states. 
\end{abstract}

\pacs{73.63.-b}
\keywords{InAs nanowires, core-shell nanowires, electron mobility, electron transport}

\maketitle 

InAs nanowires are a promising material for use in nanoscale circuits \cite{Nam09}, single electron charge sensing \cite{Shorubalko2008,Salfi2010}, and infrared optoelectronics applications \cite{Sun2010}, and due to a large spin-orbit interaction, are of potential interest for spin-based \cite{Nadj-Perge10} and topological \cite{mourik2012} quantum information processing. The intrinsic donor-like surface states of InAs play a major role in determining transport properties \cite{Dayeh2007}, and lead to reduced electron mobilities at low temperatures due to ionized impurity scattering \cite{Affentauschegg2001}. The charged surface states produce a random spatial electrostatic potential in the nanowire, which may contribute to the spontaneous formation of quantum dots at low temperature \cite{Schroer2010}. These states can also quench intrinsic photoluminescence, severely limiting the performance of optoelectronic devices \cite{Sun2010, Sun2012}. Surface passivation should reduce the density of ionized surface states and lead to improved electron mobility. Growth of an InP shell on an InAs core \cite{Tilburg2010} has been shown to yield better mobilities, as well as chemical passivation based on In-S bonding \cite{Hang2008, Petrovykh2003, Sun2012}. Here we study the effects of surface passivation due to an epitaxial In$_{0.8}$Al$_{0.2}$As shell. The bandgap of In$_{0.8}$Al$_{0.2}$As should be 0.41 eV larger than that of InAs \cite{Vurgaftman2001} for an unstrained zincblende structure, however our nanowires are wurtzite and strain is likely to modify the bandgaps somewhat \cite{Pistol2008}. We note that the lattice mismatch to InAs is smaller for In$_{0.8}$Al$_{0.2}$As shells compared to InP shells, which should result in less interfacial strain and the ability to grow thicker shells free of dislocations. We performed transport measurements on core-shell nanowire field-effect transistors (FETs) at temperatures ranging from 1-200 K. By comparing with transport results on pure InAs nanowires grown under nominally identical conditions, we obtain clear evidence for the reduction of ionized impurity scattering in the core-shell nanowires, leading to the highest mobility we have yet observed in an InAs FET device at low temperature. These results show that InAlAs epitaxial shells may offer significant improvement to the quality of nanoelectronic devices based on InAs nanowires. 

\begin{figure}
\includegraphics[width=1\textwidth]{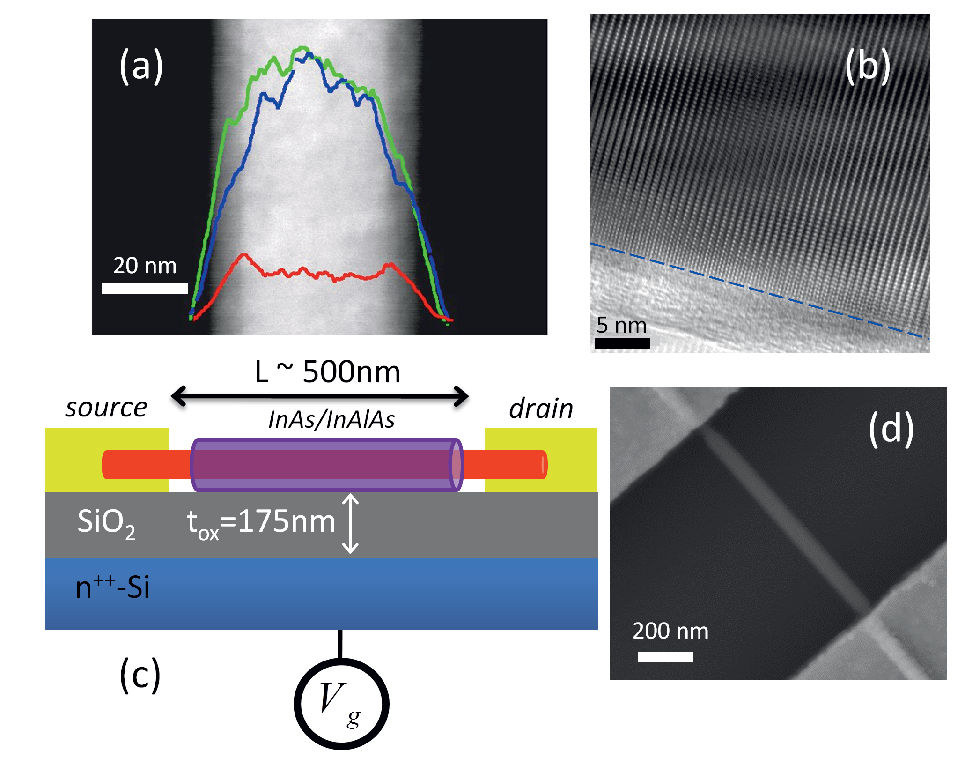}
\caption{(a) High-angle annular dark field image of InAs - In$_{0.8}$Al$_{0.2}$As core-shell nanowire with superimposed energy-dispersive x-ray spectroscopy linescan (Al: red, In: blue, As: green). (b) High-resolution TEM image taken along the $\left[2\bar{1} \bar{1}0\right]$ zone axis, showing an absence of dislocations in the shell. The dashed line indicates the nanowire surface. (c) Schematic cross-section of the FET device. (d) SEM image of device 3. The etching profile of the nanowire is seen near the metal contact at the upper left. }\label{fig1}
\end{figure}

Core-shell nanowires were grown in a gas source molecular beam epitaxy system using Au seed particles \cite{plante2009}. First, InAs cores were grown axially by the Au assisted vapour liquid solid mechanism on a p-GaAs (111)B substrate at a growth temperature of 420 $^{\circ}$C. This was followed by the inclusion of Al to facilitate radial growth of the In$_{0.8}$Al$_{0.2}$As shell \cite{Haapamaki2012}.The nanowires were characterized using transmission electron microscopy (TEM). As-grown nanowires were sonicated and suspended in ethanol, dispersed onto TEM grids with holey carbon films, and imaged with a JEOL 2010F TEM with an accelerating voltage of 200 kV. Low-magnification TEM images of typical nanowires reveal that the nanowires have an inner core and an outer shell structure. In general, the nanowires had a core diameter of 20-50 nm and a shell that was 12-15 nm thick, independent of core diameter. The chemical composition of the nanowires was analyzed by energy-dispersive x-ray spectroscopy (EDS). As shown in Figure 1a, the EDS line scan analysis along the radial direction shows In and As in the core region and In, As and Al in the shell region. High-resolution TEM (HRTEM) image of a representative unetched nanowire in Figure 1b clearly shows lattice fringes of a single-crystal nanowire along the $\left[2\bar{1} \bar{1}0\right]$ zone axis. Both core and shell exhibit wurtzite crystal structure, evidenced by ABAB... stacking, and confirmed by selected area diffraction. HRTEM and electron diffraction data are both consistent with a dislocation-free core-shell interface for these nanowires. However, at higher Al concentrations, dislocations due to relaxation of the core-shell interface are observed \cite{Haapamaki2012}. The nanowires studied here had low stacking fault densities, achieved by using sufficiently low growth rates \cite{Shtrikman2011, Haapamaki2012}. Below, we also show data from unpassivated InAs nanowires that were grown under nominally identical conditions to the growths of the nanowire cores mentioned previously, except that the axial growth rate was 0.25 $\mu$m/hr and 0.5 $\mu$m/hr for the core-shell and bare nanowires, respectively.

FET devices were fabricated using a standard e-beam lithography technique. As-grown nanowires were mechanically deposited onto a 175 nm thick SiO$_2$ layer above a n$^{++}$-Si substrate. Selected nanowires with diameters $40-80$nm were located relative to pre-existing fiducial markers by scanning electron microscopy (SEM), with care taken to minimize the electron dose. The contact areas were etched with citric acid to remove the shell material, followed by room temperature sulfur passivation to prevent oxide regrowth \cite{Suyatin2007} during the sample transfer to an e-beam metal evaporator. Ni/Au (30nm/50nm) metal contacts were deposited and annealed at 120 $^{\circ}$C in vacuum for 5 minutes to promote Ni diffusion into the nanowire contact area \cite{Chueh2008}. The device structure is shown schematically in Fig. 1c, and a SEM image of a representative device is shown in Fig. 1d. After fabrication the device chip was wire-bonded into a chip carrier and transport measurements were carried out in either a continuous flow He cryostat operating from 4-200 K or a dilution refrigerator allowed to slowly warm from 1.2 K. Upon applying a DC source-drain voltage $V_{sd}$, the device current $I_{sd}$ was measured using a current-voltage preamplifier at a 3 Hz bandwidth. A separate voltage $V_g$ applied to the degenerately doped Si substrate provided a global backgate. 

\begin{figure}
\includegraphics[width=1\textwidth]{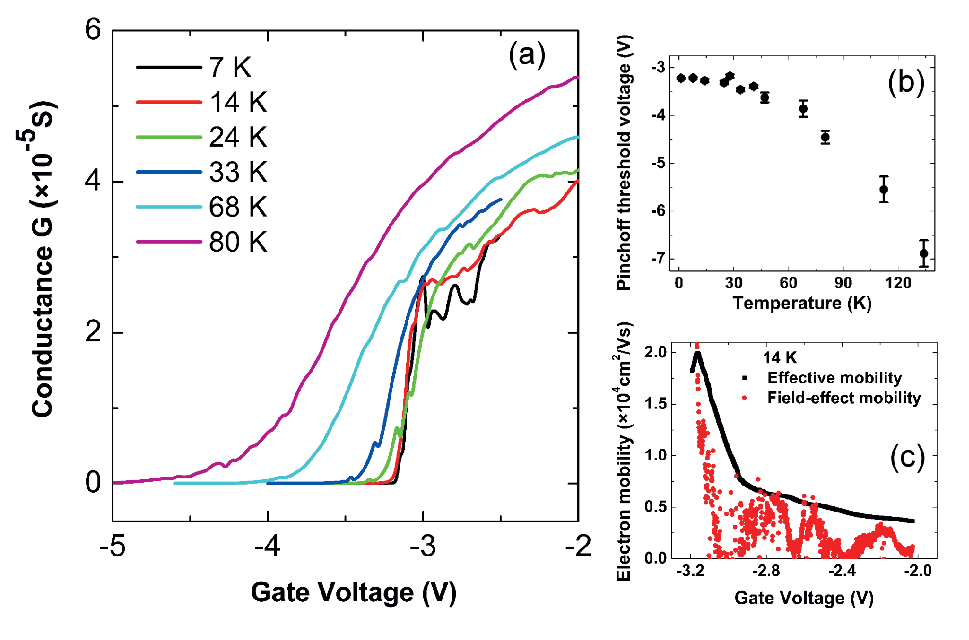}
\caption{(a) Conductance versus backgate voltage for core-shell nanowire FET device 1, at various temperatures. (b) Pinchoff threshold voltage as a function of temperature for device 1. (c) Field-effect (red dots) and effective (black squares) mobility versus backgate voltage for device 1 at $T=14$ K. The effective mobility remains a smooth function despite the oscillations in conductance that appear at low temperatures. The peak effective mobilities are used in subsequent analysis. }\label{fig2}
\end{figure}

Many devices have been investigated to different levels of detail, but here we will focus on results from three core-shell nanowire FETs (devices 1-3) and three pure InAs nanowire FETs (devices 4-6) for which detailed mobility analysis was performed. Pinchoff threshold data from additional core-shell devices is also included in Fig. 3. Devices 1-3 had total diameters 80 nm, 75 nm, and 51 nm, respectively, measured after transport measurements by SEM at a magnification of 100k, with uncertainties $\pm 2$ nm. These correspond to core diameters of roughly 54 nm, 49 nm and 25 nm, respectively. The FET channel lengths were 465 nm, 465 nm, and 940 nm, respectively. Figure 2a shows selected conductance $G=I_{sd}/V_{sd}$ versus backgate curves for device 1. As temperature is lowered, the device `on' conductance decreases due to decreasing carrier density, but the field-effect mobility increases, indicated by the increasing slope $\frac{dG}{dV_g}$ near conductance pinchoff. Pinchoff threshold voltages $V_T$, shown in Fig. 2b versus temperature, are determined by the $G=0$ intercept of the line tangent to the maximum slope in conductance. $V_T$ shifts to more negative values with increasing temperature; this can be understood by the thermal activity of surface donor states causing the downward band bending at the surface to increase, leading to a larger accumulation of carriers at the surface and requiring larger negative gate voltages to reach pinchoff. \\
\indent The field-effect and effective mobilities, $\mu_{\text{fe}}$ and $\mu_{\text{eff}}$, are deduced from the conductance data using two different analytical expressions:
\begin{align}
\mu_{\text{fe}} &= \frac{L}{C'_g}\frac{dG}{dV_g} \\
\mu_{\text{eff}}&= \frac{L}{C'_g}\frac{G}{(V_g-V_T)}
\end{align} 
Here, $L$ is the channel length, $V_g$ is the gate voltage, $V_T$ is the threshold voltage, and $C'_g$ is the gate capacitance per unit length. $C'_g$ is estimated as the capacitance of a wire above an infinite conducting plane in series with the cylindrical capacitance between the core and shell. In estimating $C'_g$, we take into account that the nanowire is not embedded in the SiO$_2$ dielectric \cite{wunnicke2006}. The field-effect mobility is strictly only valid at the peak mobility where $\frac{d\mu}{dV_g}=0$, and it is generally a lower bound on the effective mobility. Figure 2c shows the comparison of the field-effect and effective mobilities for device 1 at T = 14 K. The effective mobility is typically a smoother function of $V_g$ than the field-effect mobility, and $\mu_{\text{eff}} \geq \mu_{\text{fe}}$ within experimental error for all our data. In the following, we take the effective mobility as our preferred measure. Two regimes of $\mu_{\text{eff}}$ are seen with respect to gate voltage in Fig. 2c, with a sharp rise in mobility near pinchoff, in contrast to a slowly varying mobility at more positive gate voltages. The decrease in mobility as the device is turned on is understood by the increase in inter-subband scattering as more radial subbands are filled \cite{Das2005}, and a larger carrier density near the surface.\\
\indent In figure 3, we compare the peak effective mobilities of the core-shell and pure InAs nanowires across a wide temperature range. The unpassivated InAs devices 4-6 had diameters 71 nm, 50 nm, and 35 nm, respectively, with uncertainties $\pm 2$ nm. The channel lengths were 2.95 $\mu$m, 970 nm and 770 nm respectively. Only the unpassivated devices exhibit a turnover at low temperatures, where mobility rapidly decreases with decreasing temperature. In contrast, the mobilities of the core-shell nanowires increase monotonically as temperature is lowered, even down to $T=1$ K (effective mobility could not be properly measured below 1 K due to the onset of strong Coulomb blockade). The turnover in mobility for the unpassivated nanowires occurs between 25 and 45 K. We have investigated several more specimens of both nanowire types, and see a similar turnover in mobility only for the unpassivated nanowires. The data in Fig. 3 have been fit to empirical power law expressions to guide the eye. The fits for the unpassivated nanowires are of the form $\mu_{\text{eff}} = (\frac{1}{\alpha T^a}+\frac{1}{\beta T^b})^{-1}$, based on combining mobilities from two scattering mechanisms using Matthiesen's rule. The fits for the core-shell nanowires are single power law functions, $\mu_{\text{eff}} = \gamma T^c$. The  temperature power law exponent describing the negative slope region for the pure InAs nanowires varies from -0.5 to -1.5, whereas it varies from -0.3 to -0.7 for the core-shell devices. We note that the highest core-shell nanowire mobility observed was $\approx 25,000$  cm$^2$V$^{-1}$s$^{-1}$ at 1 K for device 1 (core diameter $\approx 54$ nm), higher than the peak mobility $\approx 20,000$ cm$^2$V$^{-1}$s$^{-1}$ at 35 K for the unpassivated device 4 (diameter $\approx 70$ nm). The latter was the highest mobility we observed in MBE-grown InAs nanowires prior to investigating the core-shell nanowires. Mobilities in excess of 20,000 cm$^2$V$^{-1}$s$^{-1}$ have also been observed with InAs-InP core-shell nanowires \cite{Tilburg2010} at low temperature. \\
\indent For completeness, we show the estimated carrier electron concentrations at peak mobility for devices 1-6 in figure 4. The concentrations are estimated from the ratio of conductivity to effective mobility. The core-shell carrier concentrations are seen to decrease steadily as temperature is lowered, whereas the unpassivated nanowire concentrations show more experimental scatter, but appear to vary less with temperature below 100 K. Note that the error bars shown only account for error in the fitting used to extract effective mobility, but there is additional experimental variation in device conductance over the course of measurements due to hysteretic changes in surface state population \cite{Dayeh2007}, trapping of electrons in the nanowire oxide \cite{Salfi2010}, etc. The core-shell concentrations fit reasonably well to a simple exponential, $n(T)=A e^{B T}$, where $n$ is concentration, $T$ is temperature, and $A,B$ are fitting parameters.

\begin{figure}
\includegraphics[width=0.8\textwidth]{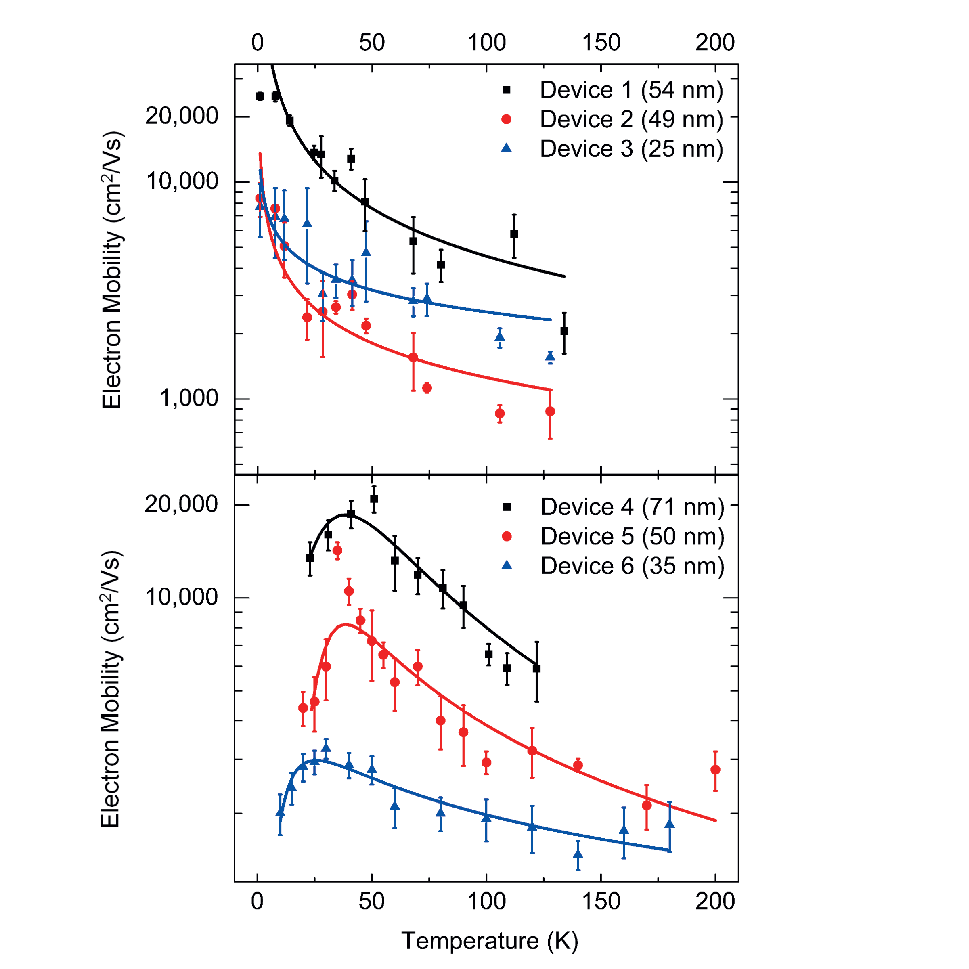}
\caption{Peak effective electron mobility versus temperature for core-shell devices 1-3 (upper panel) and unpassivated devices 4-6 (lower panel). The estimated nanowire diameters (core diameters for devices 1-3) are indicated in the legend in parentheses. Empirical fits are based on a single power law $\mu_{\text{eff}} = \gamma T^c$ for the core-shell devices, and a two-component form $\mu_{\text{eff}} = (\frac{1}{\alpha T^a}+\frac{1}{\beta T^b})^{-1}$ for the unpassivated devices, where $\alpha, \beta, \gamma, a, b, c$ are fitting parameters. On average, mobility increases with nanowire diameter, as is seen here for the unpassivated nanowires. 
}\label{fig3}
\end{figure}

\begin{figure}
\includegraphics[width=0.8\textwidth]{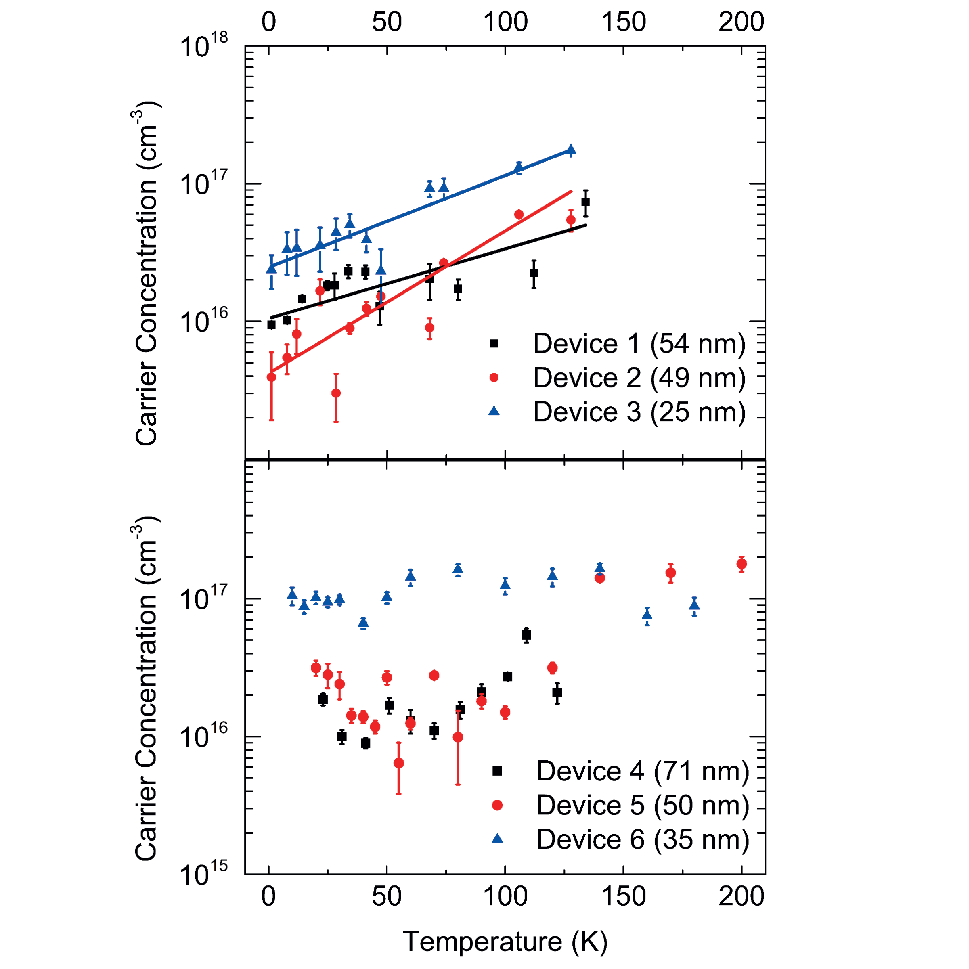}
\caption{Carrier electron concentrations at peak mobility, versus temperature, for core-shell devices (upper panel) and unpassivated devices (lower panel). The estimated nanowire diameters are indicated in the legend. The solid lines in the upper plot are empirical fits of the form $n(T)=A e^{B T}$, where $n$ is concentration, $T$ is temperature, and $A,B$ are fitting parameters. The slopes $B$ obtained are $0.012$ K$^{-1}$, $0.024$ K$^{-1}$ and $0.015$ K$^{-1}$ for devices 1, 2 and 3, respectively. The concentrations for the unpassivated nanowires show more scatter, but are largely independent of temperature below 100 K.   
}\label{fig4}
\end{figure}

\indent To explain the results for mobility, we must first consider the possible scattering mechanisms for conduction electrons in InAs nanowires. Acoustic phonon scattering can be ruled out as the limiting mechanism, because it would yield much higher mobilities $\sim 5\times 10^6$ cm$^2$V$^{-1}$s$^{-1}$ at room temperature \cite{madelung1964}. Similarly, optical phonon scattering is only expected to dominate at much higher temperatures \cite{madelung1964}. Surface roughness and ionized impurity scattering are the most plausible mechanisms. Although we cannot strictly rule out the presence of bulk impurities, we expect their densities to be very low in comparison to the density of surface states. Surface roughness scattering should only limit the mobility associated with a surface accumulation layer, as mobilities in thin quantum wells have been shown to scale with the well width to the sixth power \cite{Sakaki1987}, and only be limited to the mobility values we observe for sub-10 nm thick conduction layers. Moreover, the temperature dependence of mobility for surface roughness scattering is expected to be weak or even to increase with temperature \cite{Sakaki1987}. Since peak mobility occurs close to pinchoff, where we expect little or no contribution to effective mobility from a surface accumulation layer, we rule out this mechanism to explain the data of figure 3. On the other hand, surface roughness scattering could well play a role away from peak mobility where the device is fully `on' at more positive gate voltages (see figure 2c). The presence of the donor-like surface states \cite{Dayeh2007, Hang2008} at a density of $10^{11}-10^{12}$ cm$^{-2}$, together with the large surface-to-volume ratio inherent to the nanowire geometry, suggests that Coulombic scattering at the surface is most likely dominant. This is consistent with the diameter dependence seen in Fig. 3; the mobility generally increases with diameter both for the core-shell and unpassivated nanowire devices, similar to previous reports \cite{Ford2009}.\\
\indent The downturn in mobility below $\sim$ 50 K for the pure InAs nanowires is consistent with ionized impurity scattering \cite{madelung1964}; the scattering rate increases as the average carrier energy decreases upon lowering temperature. On the other hand, how can we understand the negative slope of $\mu_{\text{eff}}$ versus $T$ for the core-shell nanowires, and for the pure nanowires above 50 K? We suggest this can be understood by the change in surface state scattering rate as both the ionization of surface states and the concentration of conduction electrons change with temperature. Thermally activated surface state ionization yields an increase in concentrations of both scatterers and carriers with temperature. At low temperature, for conduction electrons in the core, the lowest radial subband is primarily occupied. This wavefunction has a relatively small overlap with the surface, and there are fewer ionized surface states, leading to a relatively high mobility. As temperature increases, higher radial subbands are occupied, leading to a greater fraction of electrons near the surface, enhanced by the negative surface band bending \cite{Olsson1996}. More electronic states become available for intersubband scattering \cite{Das2005}, and together with more thermally activated scatterers, the average mobility decreases. The behaviour of mobility and carrier concentration below 50 K in the unpassivated nanowires suggests a temperature independent density of charged surface scatterers, in contrast to the core-shell nanowire data. A fixed density of scatterers would indeed be expected to produce a positive slope of mobility versus temperature, since $\mu \propto E^{3/2} \propto T^{3/2}$, where $E$ is the carrier energy \cite{madelung1964}. One also expects a $T^{3/2}$ temperature dependence for carrier concentration in three dimensions, however, a weaker $T^{1/2}$ dependence arises in one dimension, which may be appropriate to describe the present system at low temperature and close to pinchoff. Although the estimated carrier concentrations for the unpassivated devices do not appear to match this expectation, we consider the data too imprecise to draw any definite conclusions. The mobility data, on the other hand, clearly suggests a reduction in the density of scatterers in the passivated nanowires at low temperature. It is not clear whether the surface passivation changes the nature of the surface states at the InAs/InAlAs interface, or simply reduces their density. An interesting possibility is that the primary electron donors in the passivated nanowires may be surface state of the outer shell, rather than the InAs/InAlAs interface. In that case, the InAs/InAlAs interface could have a reduced density of surface states, and the high density of outer shell surface states would be much less effective in scattering conduction electrons in the core. \\
\indent In conclusion, we have shown evidence through electron transport for the effectiveness of an In$_{0.8}$Al$_{0.2}$As shell to passivate the InAs nanowire surface. The marked difference in low temperature mobility between the passivated and unpassivated nanowires suggests a reduction in charged InAs surface states by the core-shell passivation. Whether the passivation changes the nature of the surface states or only reduces their number is yet to be determined. Regardless, these core-shell nanowires are expected to offer new opportunities to realize high quality devices for quantum transport and optoelectronics applications. \\

We would like to acknowledge the Canadian Centre for Electron Microscopy, the Centre for Emerging Device Technologies, and the Quantum NanoFab facility for technical support. We thank Shahram Tavakoli for assistance with MBE and Roberto Romero for general technical assistance. Helpful discussion with Prof. Xuedong Hu is also acknowledged. This research was supported by NSERC, the Ontario Ministry of Economic Innovation and the Canada Foundation for Innovation. G. W. H. acknowledges a WIN Fellowship.

%

\end{document}